\documentstyle[prl,aps,epsfig,floats]{revtex}

\begin{document}
\draft

\textheight=23.8cm
\twocolumn[\hsize\textwidth\columnwidth\hsize\csname@twocolumnfalse\endcsname

\title{\Large {\bf Dynamic Oscillations of Magnetization in High Spin Magnetic 
Clusters}}
\author{\bf Indranil Rudra and S. Ramasesha}
\address{\it Solid State and Structural Chemistry Unit \\ 
Indian Institute of Science, Bangalore 560012, India}

\date{\today}
\maketitle


\begin{abstract}
\noindent We have studied the time evolution of magnetization of a high spin 
magnetic cluster in the ground state, in the presence of a sinusoidal axial
magnetic field and a static transverse field by explicitly solving time 
dependent schr\"odinger equation. We observe oscillations of magnetization in 
the plateau region which has all the characteristics reminiscent of the 
oscillation of probability of occupation of a state in a two level lattice in 
the presence of an oscillating electric field. The high spin of the ground 
state leads to a large but finite two-level pseudo-lattice. The oscillations 
in magnetization occur at amplitude of magnetic fields achievable in a 
laboratory and they also persist for a wide range of spin-dipolar interactions.
 Thus, our study provides a magnetic analog of the optical two-level lattice.\\
\end{abstract}
\vskip .5 true cm

\vskip.5pc
]
\vskip .5 true cm

In recent years, synthesis of high nuclearity transition metal complexes
in high-spin ground state has spurred interest in magnetism on a nanoscale\cite
{synth}.
The synthesis of Mn$_{12}$ and Fe$_8$ clusters in S=10 ground state has 
led to extensive study of quantum resonant tunneling and quantum interference
phenomena\cite{qtm}. The quantum resonant tunneling manifests as plateaus in the
magnetization {\it vs} magnetic field plots, with the width and location
of plateaus being determined by the ramping speed of the magnetic field as 
well as the initial state\cite{rudra}. The quantum interference phenomena 
observed in the $\rm Fe_8$ cluster is because the paths connecting the $M_S=+10$
and $M_S=-10$ could interfere in the presence of a magnetic field, leading to
an oscillation in the tunneling probabilities\cite{wernsdorfer}.

In these systems, the magnetic interaction between transition metal ions in 
neighbouring molecules is weak due to the bulky nature of the molecular 
complex. Thus, we can treat these molecules as isolated magnetic clusters. 
These clusters are characterized by exchange couplings between magnetic ions 
that are frustrated, leading to a high spin ground state separated from other 
low-lying high spin states by rather small energy gaps. These magnetic clusters
also exhibit signatures of spin dipolar and other higher order spin-spin 
interactions allowed by the symmetry of the cluster.

The exchange Hamiltonian of the Mn$_{12}$ system with which we shall be 
concerned in this letter can be written as
\begin{eqnarray}
{\rm
  {\hat H_{exch}} ~ = ~\sum_{<ij>}J_{ij} {\hat s}_i \cdot {\hat s}_j}
\end{eqnarray}
where the exchange interactions are given in fig.1. The model exact S=10 
ground state as well as the low-lying excited states of this Hamiltonian have 
been obtained from this model by using all the spatial symmetries and partial
spin symmetry adaptation. The ground state and the first few low-lying
states live in different symmetry subspaces and do not mix unless the molecular
symmetry is broken\cite{raghu}. Therefore, for computing properties such as 
quantum resonant tunneling at low-temperatures, it is sufficient to work with 
the different $M_S$ manifolds within the $S=10$ ground state. The spin-spin
interactions allowed by the symmetry of the molecule leads to a spin
Hamiltonian in the presence of an applied magnetic field given by
\begin{eqnarray}
{\rm
\hat{H}_{dip} =  D~\hat{S}_{z,total}^2 + c~(\hat{S}_{x,total}^4 +
\hat{S}_{y,total}^4)} \nonumber \\
{\rm
+~ {\it g}_{crown}~{\vec H} (t) \cdot \hat{S}_{crown} +
{\it g}_{core }~ {\vec H} (t) \cdot \hat{S}_{core}~}
\end{eqnarray}
where, for chemical reasons, the $g$ factors for the Mn$^{3+}$ and Mn$^{4+}$ 
ions in the crown and core of the molecule (fig 1) are slightly different
\cite{barra}. A time-dependent solution of the above Hamiltonian, starting from
an initial state with $S=10$ and $M_S$=-10 shows quantum resonant tunneling when
the magnetic field is ramped.\cite{rudra}

The energy level structure of the eigenstates of eqn. 2 are very similar to
those encountered in the dynamic studies of two-level lattices. There has
been considerable study of the two-level lattices in which Stark-Wannier
effect and Bloch oscillations are observed\cite{wannier,bloch}. In the presence
 of oscillatory electric fields, population trapping\cite{agarwal} and dynamic 
localization \cite{dunlap} have also been observed. The aim of the present 
study is to see if these effects can also be observed in the Mn$_{12}$ 
magnetic cluster in the presence of an oscillatory magnetic field.

In the present work we have studied the magnetization oscillations under the
influence of a strong axial {\it ac} magnetic field and a weak stationary
transverse field, using explicit time evolution of an initial state. 
We start with an initial state which is the lowest eigenstate of the full 
exchange Hamiltonian (eqn. 1) with $S=10$ and $M_S=-10$. In the absence 
of the quartic spin terms and external transverse field, this is also the 
lowest energy state of the low-energy Hamiltonian (eqn. 2). We evolve this 
state with time explicitly, using the time-dependent schr\"odinger equation,
\begin{equation}
{\rm
i \hbar {\frac{\partial \psi(t)}{\partial t}}=H_{dip.}\psi(t)}.
\end{equation}
At each time step, we calculate the magnetization of the state. The value 
of D estimated from experiments is 0.56K, while c is $3 \times 10^{-5}$K \cite
{barra}. The $g_{\rm crown}$ and $g_{\rm core}$ are respectively 1.96 and 2.00.
In our studies, we have varied the amplitude of the time dependent magnetic 
field between 10 to 100 D. These fields range from 3.75 T to 37.5 T , for 
experimentally estimated D value in Mn$_{12}$ , which are indeed attainable 
\cite{field}. The transverse field is static and fixed at a value equal to one 
hundredth the amplitude of the time varying axial magnetic field.

The initial state in the time evolution is chosen to be the state with 
$S=10$ and $M_S$ = -10, which is the ground state, in the absence of the
weak off-diagonal terms of H$_{\rm dip}$. The angular frequency, $\omega$, 
of the axial field is varied between 10$^{-1}$ and 10$^{-3}~
{\rm radian-D/\hbar}$. The time
evolution is carried out successively in steps $\Delta t=0.1$ and the 
evolution is carried out over several periods of the applied field.

In fig. 2a, we show a plot of magnetization {\it vs} time, for the amplitude
of the axial field, H$_0$=50D, and angular frequency $\omega= 10^{-2}~
{\rm radian-D/\hbar}$.
We notice that the magnetization shows distinct plateaus and in each
plateau, the magnetization oscillates a fixed number of times. The number
of oscillations N$_{\rm osc.}$ in each plateau is given by the ratio of half
the energy gap between the $M_S =-10$ and the $M_S=-9$ states at the field 
H$_0$ and $ \hbar \omega$, where $\omega$ corresponds to the angular
frequency of the applied axial field. This kind of oscillations with similar
dependence of N$_{\rm osc.}$ has been theoretically observed in the two-level
systems by Rotvig $et~al$\cite{rotvig} and Raghavan $et~al$\cite{raghavan}. 
Such oscillations have also been seen in the Bose-Einstein condensates in a
double-well trap\cite{smerzi}. Rotvig $et~al$ observe this in the context of a 
two-band semiconductor superlattice in an external electric field, while 
Raghavan $et~al$ observe it in a single band model in the presence of an 
electric field. In these calculations, the probability in a given state shows 
the temporal oscillations that we see here for magnetization. Indeed, the 
different $M_S$ states in the magnetic cluster can be viewed as forming a large
but finite lattice. The transverse magnetic field couples the states at 
successive lattice sites, much as the transfer terms in the models of Rotvig 
$et~al$ and Raghavan $et~al$. The time varying axial magnetic field corresponds
to the applied oscillatory electric field of the two level systems. It is also 
worth noting that for the parameters that Raghavan $et~al$ use, the 
oscillations die down for larger lattices. It appears that the size of the 
pseudo-lattice provided by the magnetic cluster is not large enough for the 
oscillations to die down for the realistic model parameters we have chosen.

The oscillations in magnetic field that we observe are quite robust. We have
observed these oscillations for other initial states with integral $M_S$ 
values corresponding to the eigenstates of the H$_{\rm dip.}$, with 
off-diagonal elements set to zero (fig. 2 b-d). The dependence of the 
magnetization on the amplitude of the magnetic field is shown in fig.3. We note
that at higher fields, we see more oscillations in each plateau since the gap 
between successive $M_S$ states widen with increasing amplitude. In fig. 4 b,d 
and f, we show the energies of the two low-lying eigenstates of H$_{\rm dip.}$, 
corresponding to the axial field at that instance. We note that the jump in 
magnetization coincides with two states with different $M_S$ values becoming 
degenerate. In the time between these coincidences, the magnetization shows 
small amplitude oscillations. Assuming that the system wavefunction evolves as
exp(-i$\rm{(\bar{E}_2-\bar{E}_1)t/\hbar}$), where ${\rm \bar{E}_1}$ and 
${\rm \bar{E}_2}$ are the average energies of the two states in question, the 
number of oscillation in a time period ${\rm t=2\pi/\omega}$ is given by
${\rm (\bar{E}_2-\bar{E}_1)/\hbar \omega}$. We note that the energy gap at 
the maximum amplitude in all the cases are the same and hence the number of 
oscillations in a plateau is inversely proportional to the frequency of the 
axial field. In fig. 4 a,c and e we show the dependence of the oscillation 
pattern on the frequency of the axial field with a fixed amplitude. We note 
that at decreasing frequencies, the number of oscillations in each plateau 
increases and the plateau structure itself vanishes with the magnetization 
following the magnetic field for higher frequency of the axial field. 

We have also studied this system in the presence of an axial magnetic field 
with two different frequencies, given by H$_0$cos($\omega_1t$)cos($\omega_2t$).
This leads to a beat pattern involving the sum and difference of the two 
frequencies. In fig. 5, we show the time variation of magnetization in
the presence of two different frequency magnetic fields. We note that one
set of oscillations correspond to N$_{\rm osc.}$=$\Delta E / 2~\hbar(\omega_1+
\omega_2)$ while another set of oscillations have N$_{\rm osc.}$=$\Delta E / 2~
\hbar(\omega_1 - \omega_2)$, where $\Delta E$ is the difference in energy 
between the $M_S$=-10 and  $M_S$=-9 states when the axial field is equal to 
the amplitude.

The parameters for which we have carried out the calculations corresponds to 
Mn$_{12}$. However, systems such as Fe$_8$ also have high spin ground state
although the D value is different. In order to study the effect of change
in D value on the oscillations, we have carried out these calculations for 
several H$_0$/D values. in fig. 6, we show the results of our calculations for
H$_0$/D = 20 and H$_0$/D = 10. The oscillations vanish (H$_0$/D = 20) and we 
have a M {\it vs} t behaviour that does not have much structure for this larger
H$_0$/D value. However for lower H$_0$/D value, we find that the oscillations 
persist, but with much reduced amplitude. Thus, it is possible that in other 
high spin systems, these oscillations are found at different field amplitude.

To conclude, the magnetic clusters in high spin ground state behave like 
a finite lattice of two level systems and in the presence of an oscillating
axial field and static transverse field lead to predictable oscillations 
in the plateau regions of magnetization {\it vs} time plots. The field at which
these oscillations can be observed fall in the range of 4 to 40 T with a 
frequency ranging from 1 to 10 MHz for the known high-spin magnetic clusters. 
While such behaviour is predicted in optical two level systems, their 
observation may be easier in high nuclearity high spin magnetic systems. 

\vskip 1 true cm
\leftline{\bf Acknowledgments}
\vskip .5 true cm

We thank Prof. Diptiman Sen for many helpful discussions, and the
Council of Scientific and Industrial Research, India for their grant No.
01(1595)/99/EMR-II.

\noindent {\bf Figure Captions}
\vskip 1 true cm

\noindent {1.} A schematic diagram of the exchange interactions between
the Mn ions in the Mn$_{12}$Ac molecule. The interactions $J_1 = 215 $K
and $J_2 = J_3 =86$K are antiferromagnetic, while $J_4 = -64.5 $K is
ferromagnetic. The Mn$^{4+}$ ions form the core while the Mn$^{3+}$
ions form the crown.\cite{raghu}

\noindent {2.}(a) Plot of evolution of magnetization with time (in units of $\hbar$/
D $\times$ 10$^{-1}$) for H$_0$/D =50 and $\omega$=0.01 ${\rm radian-D/\hbar}$,
initial state is $M_s$=-10.  For (b), (c) and (d) H$_0$/D =100 and 
$\omega$=0.01 ${\rm radian-D/\hbar}$, initial states are respectively 
$M_s$=-9, -4 and 4.

\noindent {3.} (a) Plot of magnetization $\it vs$ time when H$_0$/D =100 and 
$\omega$=0.01. For (b) H$_0$/D =200 and $\omega$=0.01. Initial state in both 
cases is $M_s$=-10. $\omega$ and time are in units of fig 2.

\noindent {4.} Magnetization $\it vs$ time (a, c and e) and energy of $M_s$=-10 
(solid line) and -9 (broken line) $\it vs$ time (b, d and f) for (i) 
$\omega$=0.001 (a, b) (ii) $\omega$=0.01 (c, d) and (iii) $\omega$=0.1 
(e, f). H$_0$/D = 100 for all cases. $\omega$ and time are in units of fig 2.

\noindent {5.} Plot of magnetization $\it vs$ time when H$_0$/D =100,
$\omega_1$=0.01 and $\omega_2$=0.005. $\omega$ and time are in units of fig 2.

\noindent {6.} Evolution of magnetization with time with different H$_0$/D
values, $\omega$ is kept fixed at 0.01. H$_0$/D=100, 20 and 10 respectively 
for (a), (b) and (c). $\omega$ and time are in units of fig 2.
 
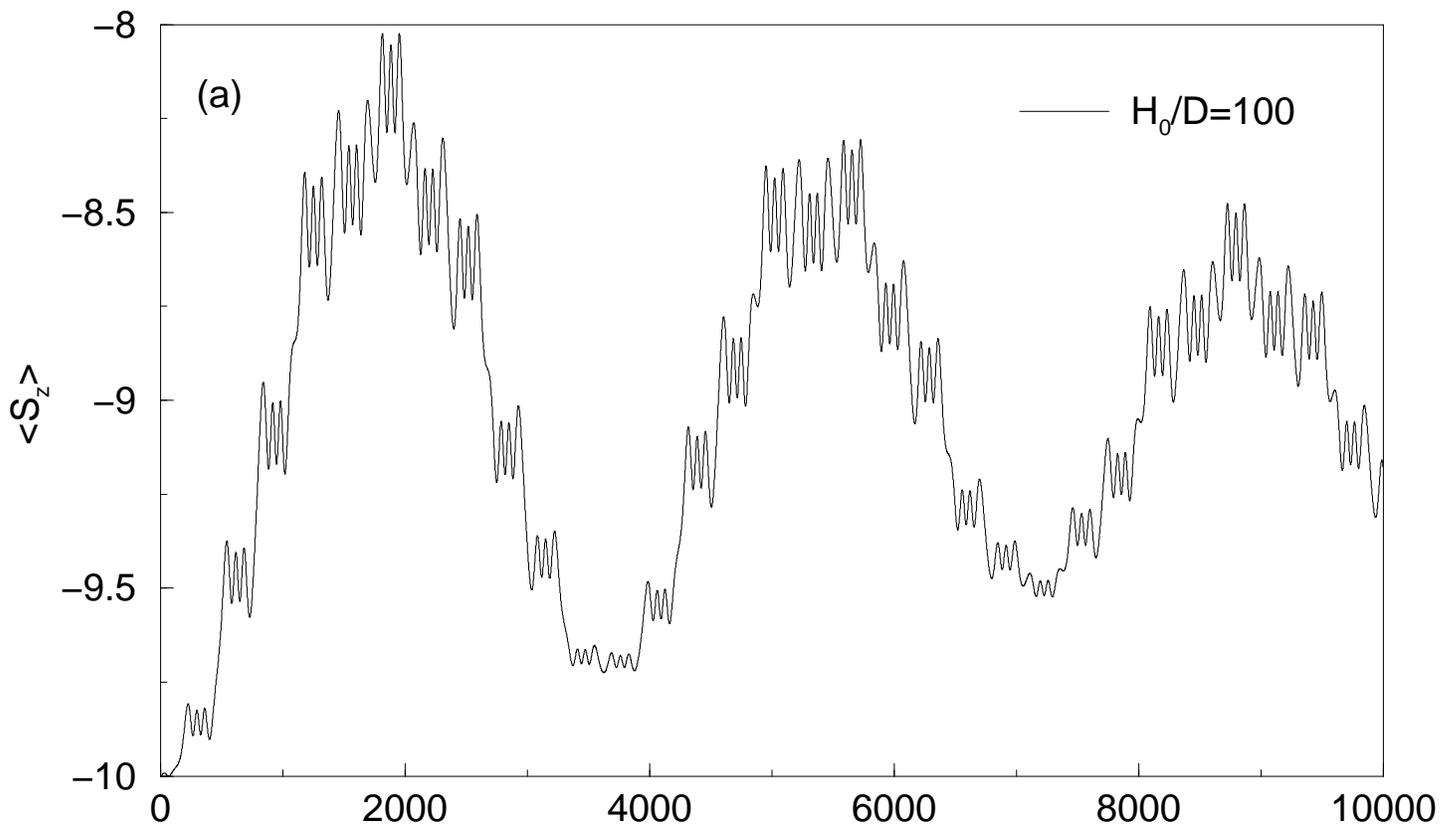
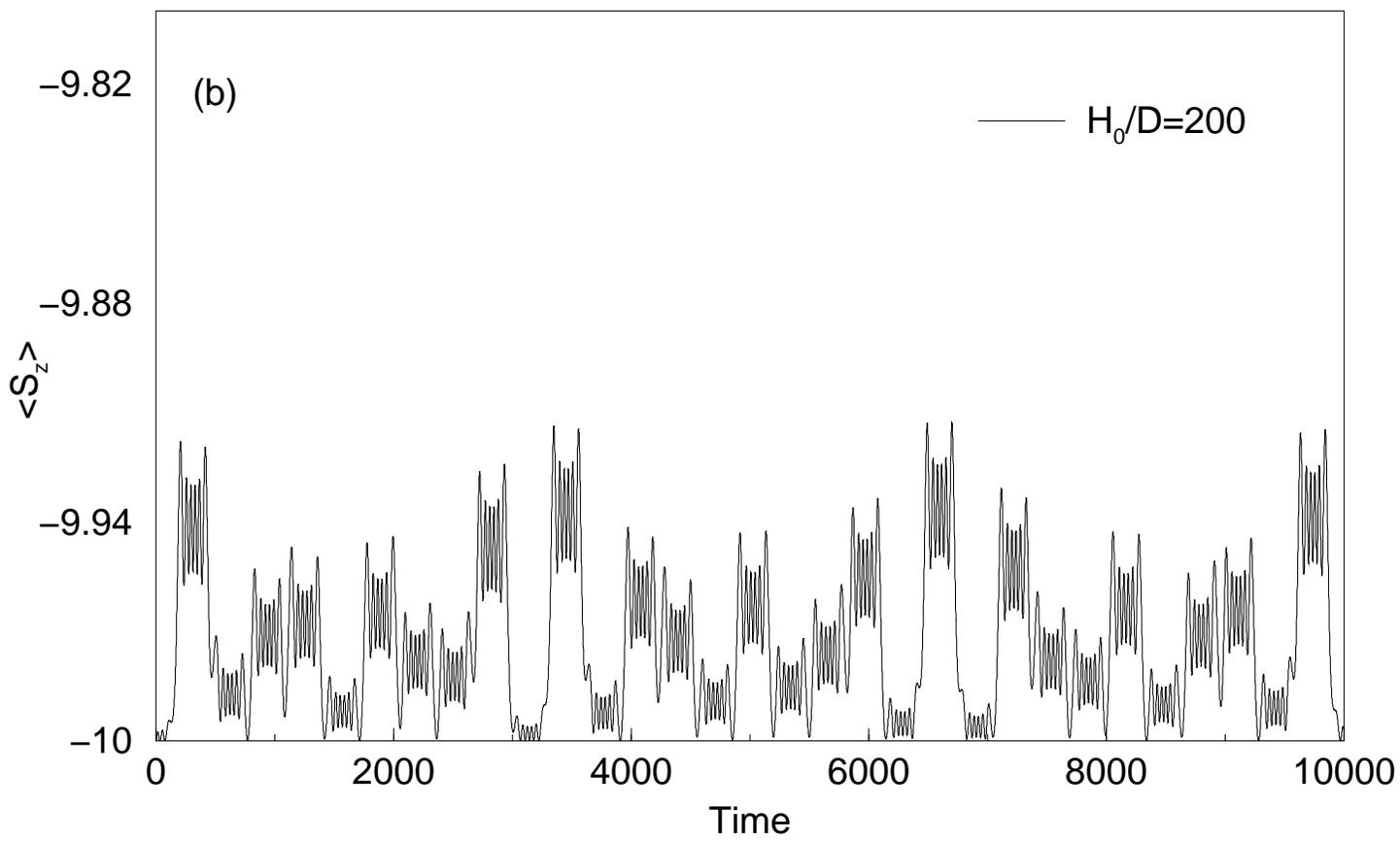
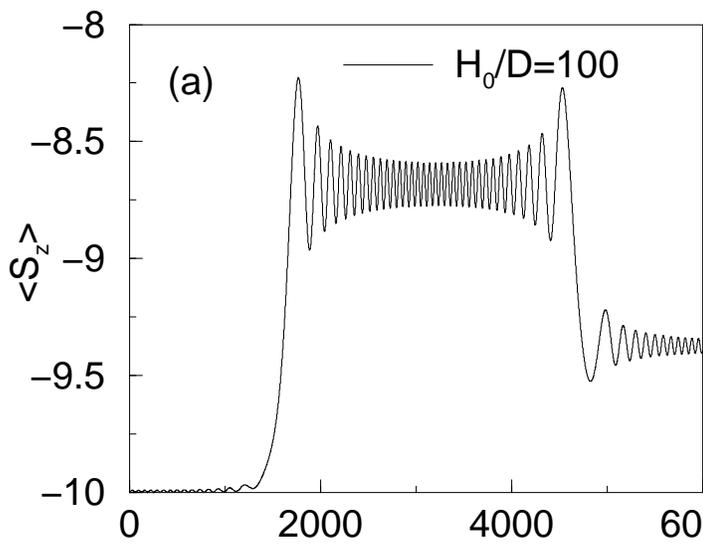
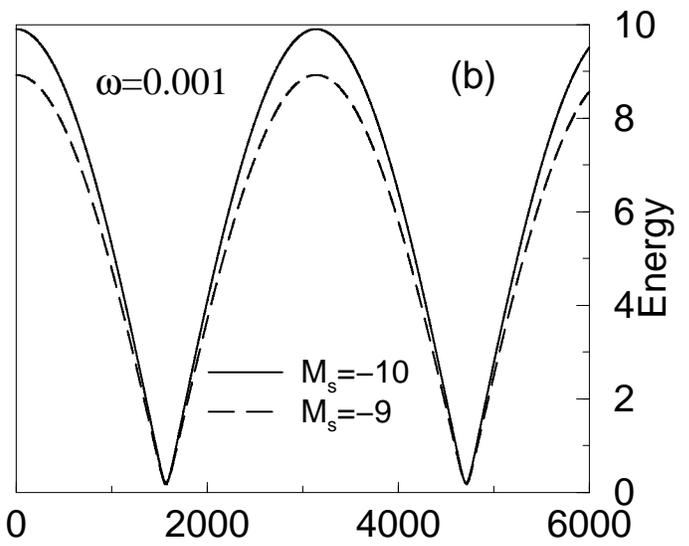
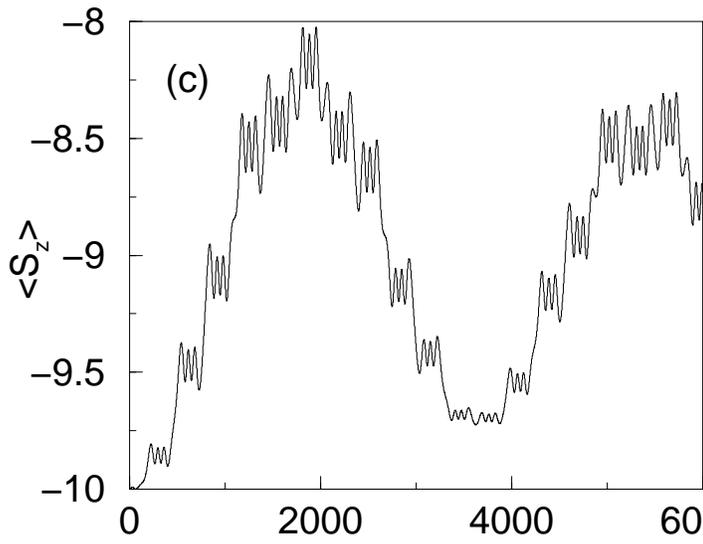
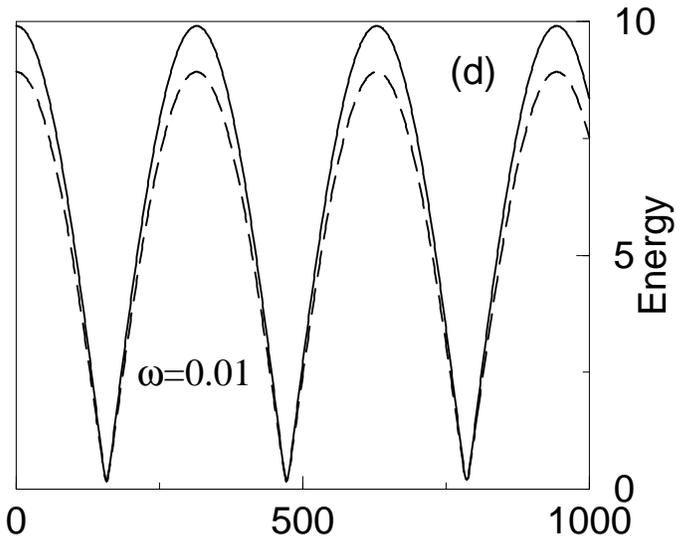
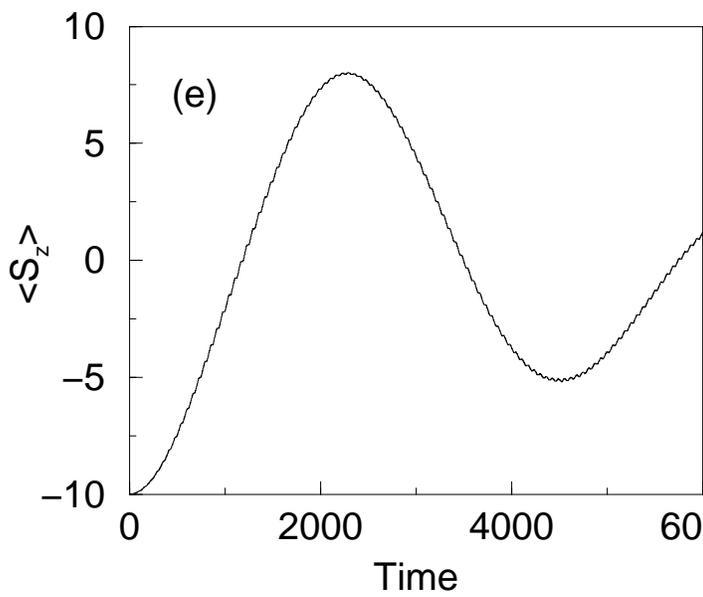
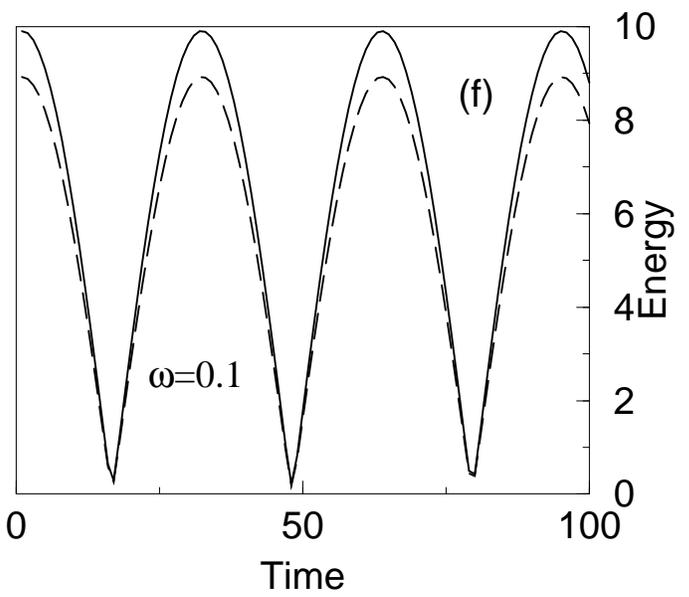
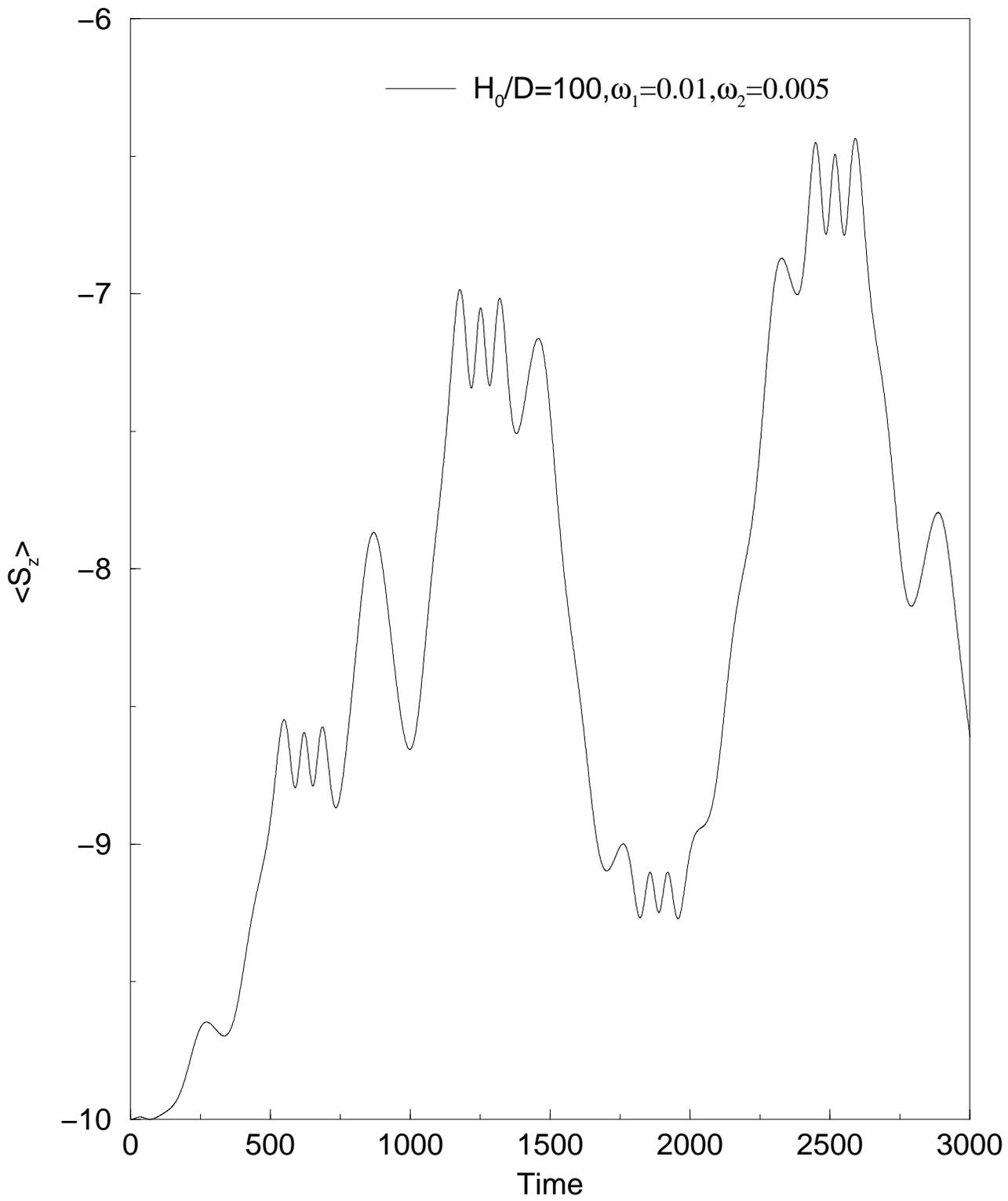
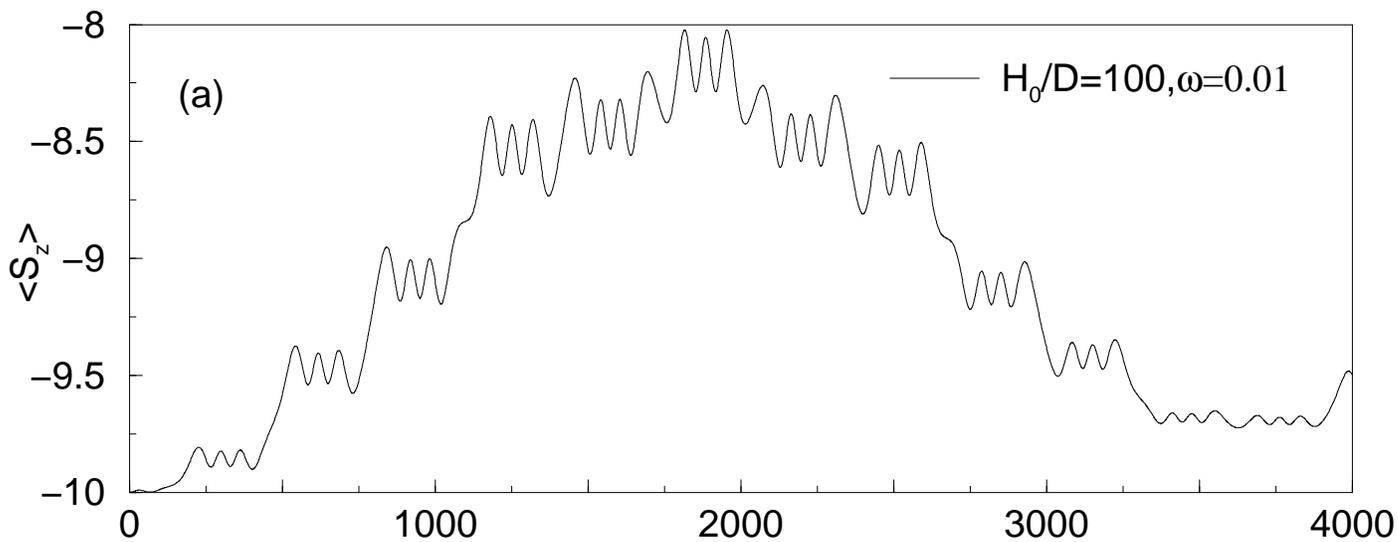
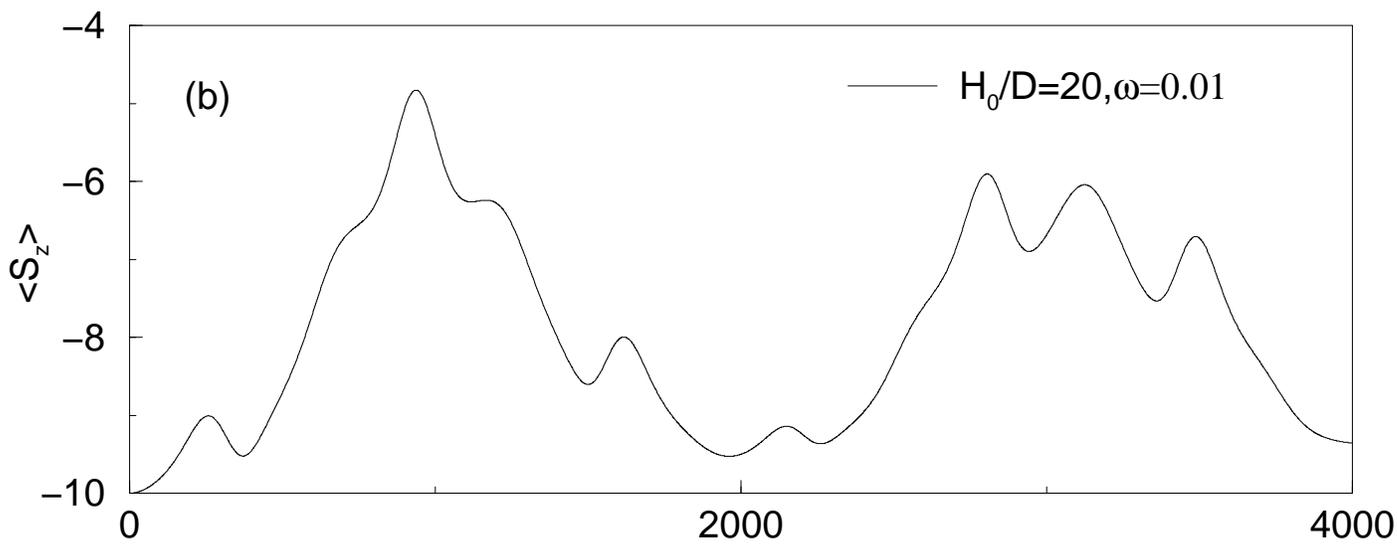
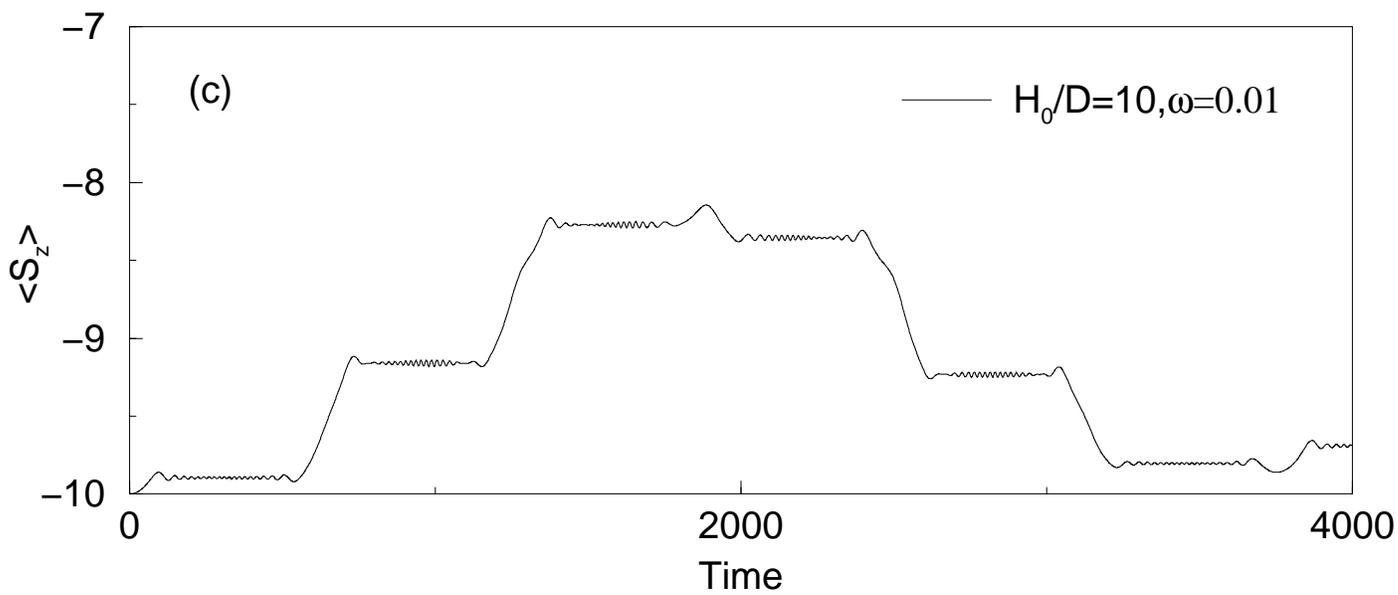
\end{document}